# Separating Current from Potential Sweep Method of Electrochemical Kinetics for Supercapacitors

Yitao He

*Room 4-304, No.5 Meishan Road, Hefei City, Anhui Province, P. R. China.*

Email: electro_yitao@163.com

**Abstract**

We briefly elaborate the concept of electrochemical kinetics firstly, and introduce a method to separate current from potential sweep method of electrochemical kinetics for supercapacitors. The current in CV curves has been separated by equating the intercepts of the equation $i_{total} = Cv + i_p$ for EDLC, and by equating the slopes of the equation for pseudocapacitance.

In the last decade, as a new energy storage device, the supercapacitors has received a great deal of attention of many researchers.[1] The most important advantage of supercapacitors is their higher power density, which making the supercapacitor grow to become an almost indispensable resource for power system.[2] The research field of the supercapacitors is an interdisciplinary field that includes the fields of electrochemistry, materials science, condensed matter physics and so on. Among them, the field of electrochemistry lays the theoretical foundation for supercapacitors. Therefore, electrochemical kinetics, as a branch of electrochemistry, plays a crucial role in the development of supercapacitors. Potential sweep method of electrochemical kinetics is found to be the most effective method to analyze and evaluate the performance of supercapacitors. Hence, in this paper, we briefly elaborate the concept of electrochemical kinetics firstly, and introduce a method to separate current from potential sweep method for supercapacitors.

**Electrochemical Kinetics and Potential Sweep Method**

Electrochemical kinetics is a field of electrochemistry studying the rate of electrochemical processes.[3] The electrochemical processes involve two elements: electro- and chemical reaction (e.g., $O \rightarrow R$, $O$ oxidizing agent; $R$ reductant). An external voltage is applied to this chemical reaction to form an electrolytic cell, in another case, a primary cell can be formed when a chemical reaction generates a voltage. The current exists in both of the electrolytic cell and the primary cell. The presence of a current represents the transferring of electrons and, therefore, indicates the presence of electron-transfer reaction, which is called electrochemical reaction (e.g., $O + ne^- \rightarrow R$). Eletrochemical reactions can be divided into three types: irreversible reaction, quasi-reversible reaction and reversible reaction. Hence, the rate in the definition of electrochemical kinetics is the electrochemical reaction rates.

As the rate ($v$) is a parameter that related to the time ($t$), therefore the kinetics must be related to $t$. In an electrochemical system, oxidizing agent/reductant ions transfer at the interface between electrode surface and eletrolyte. As we known, the ion diffusion process takes time, thus as a result the electrochemical kinetics is a theoretical science based on ion diffusion. Since Fick's law is important in diffusion theory, many formulas have been derived based on solutions of Fick's equation. If we introduce any other boundary conditions, the general electrochemical equations (i.e., *i-E* equations) could be obtained. The relationships between Fick' law, boundary conditions and applications are shown in Figure 1.

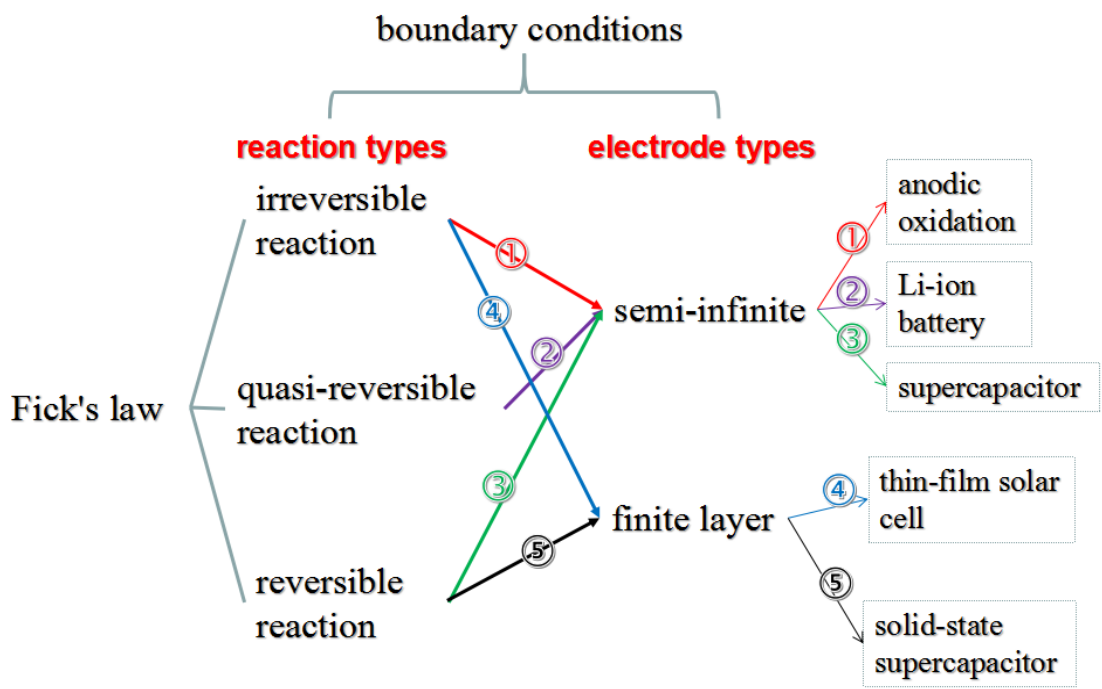

Figure 1. The classification diagram of electrochemical kinetics.

There can also have barrier layer in the electrode types, but such case is not considered here. The systems considered in this paper may be regarded as a model for the reversible reaction on a semi-infinite surface. A semi-infinite surface is bounded in one direction, and unbounded in another.

Potential sweep method is used to analyze the relationships between current, potential and time. The potential between the working electrode and the reference electrode is swept linearly in time. Therefore, the method is an important analytical tool for electrochemical system, such as the use in supercapacitors. However, since it is impossible to obtain exact *i-E* equation under the conditions of finite layer and reversible reaction, the liquid-/solid-state supercapacitors have been regarded as a system under conditions of semi-infinite and reversible reaction. Because the former math problems have not been resolved. Hence, we can get the *i-E* equation following the steps below:

Fick's law expression for the ion diffusion on the electrode surface:

$$\frac{\partial C_o(x,t)}{\partial t} = D\frac{\partial^2 C_o(x,t)}{\partial x^2} \quad (1)$$

Combining the boundary condition for reversible reaction, i.e., Nernst equation, and semi-infinite:

$$\lim_{x\to\infty} C_O(x,t) = C_o^b, \lim_{x\to\infty} C_R(x,t) = 0 \quad (2)$$

where $C_o$ and $C_R$ represent the concentrations of oxidizing agent and reductant, respectively. $C_o^b$ represents the concentration of oxidizing agent for large distance from the electrode surface.

Finally, by combining the boundary condition of potential sweep $E(t) = E - vt$, Nicholson[4] obtained an integral equation relating current to time. The exact analytical solution of this integral equation, however, has not been found. Fortunately, the peak current-scan rate ($i_{p,peak} - v$) equation can be obtained by numerical solutions:

$$i_{p,peak} = 0.4463nFAC_o\sqrt{\frac{nFD_o}{RT}}\sqrt{v} \quad (3)$$

## Separating Current

Undoubtedly, a non-ideal supercapacitor electrode could provide not only pseudocapacitance but also electric double layer capacitance (EDLC). In fact, nowadays there are many situations in which researchers must calculate the current provided by pseudocapacitance and EDLC, respectively. With the above knowledge, the current in cyclic voltammetry (CV) curves derived from potential sweep method can be easily separated.

According to the fact that a non-ideal supercapacitor electrode could provide pseudocapacitance current and EDLC current at the same time, we can get:

$$i_{total} = i_p + i_d \quad (4)$$

where $i_p$ represents the pseudocapacitance current for any potential. In Eq.(4), the value of EDLC current is:

$$i_d = C\frac{E}{t} = Cv \quad (5)$$

where $C$ represents the EDLC, which is constant ideally. Similarly, the $i_p$ is proportional to $\sqrt{v}$ at non-peak potential, and the total current value is:

$$i_{total} = A\sqrt{v} + Cv \quad (6)$$

At the peak potential, the value of A is

$$A = 0.4463nFAC_o\sqrt{\frac{nFD_o}{RT}} \quad (7)$$

*EDLC electrodes* A is a constant for an electrochemical system at a certain potential. Therefore, according to Eq.(6), plot of the $i_{total}$ as function of the $v$ illustrates their proportional relationship. We can obtain $i_p$ values by equating the intercepts of the straight-lines shown in

Figure 2.[5]

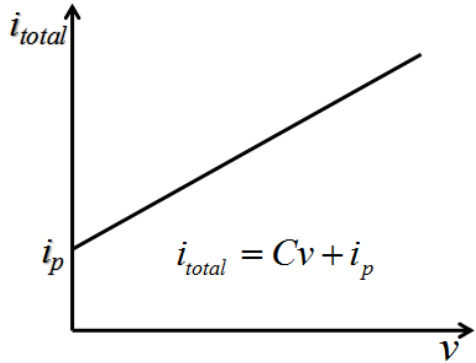

Figure 2. Plot of the $i_{total}$ as function of the $v$; the intercepts of the straight-lines is $i_p$.

In addition, the potential sweep rate exerts more influence on $i_d$ than $i_p$ when increasing $v$. Therefore, $i_d$ is not considered as a constant. We cannot obtain $i_d$ values from $i_{total} = A\sqrt{v} + i_d$ by equating the intercepts of the straight-lines. This method is only valid for the CV curves of EDLC. By applying the method in Figure 2 to any experimental CV curve, we can draw the separated EDLC current curve, and the flow-process diagram of separating current is shown in Figure 3.

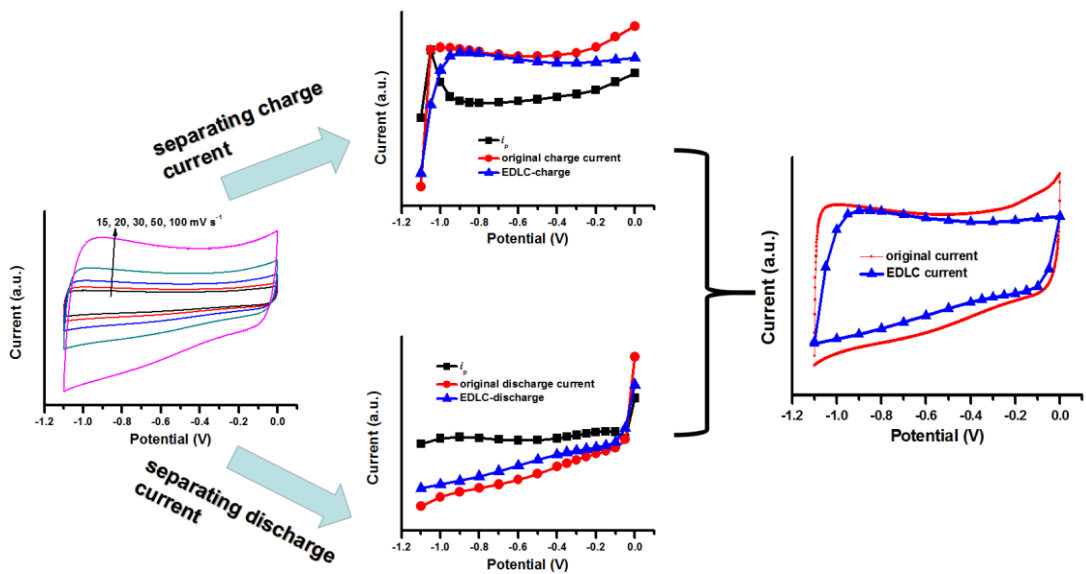

Figure 3. Flow-process diagram of separating current in CV curve of EDLC.

In Figure 3, many CV curves at different rates must be firstly measured by experiments, and the the charge current should be distinguished from the discharge current. Next, after obtaining the values of $i_{total}$ at several potentials, we calculate a series of values $i_p$ through the method in Figure 2. Finally, according to $i_d = i_{total} - i_p$, the separated EDLC CV curves are successfully obtained. Moreover, it can be seen from Figure 3 that the pseudocapacitance current exists mainly at higher potentials when charging, and at lower potentials when discharging. Therefore, as shown in Figure 4, it is seen that a chronopotentiometric (CP) curve deviates considerably from a straight line, over the existing of the potential range of pseudocapacitance.

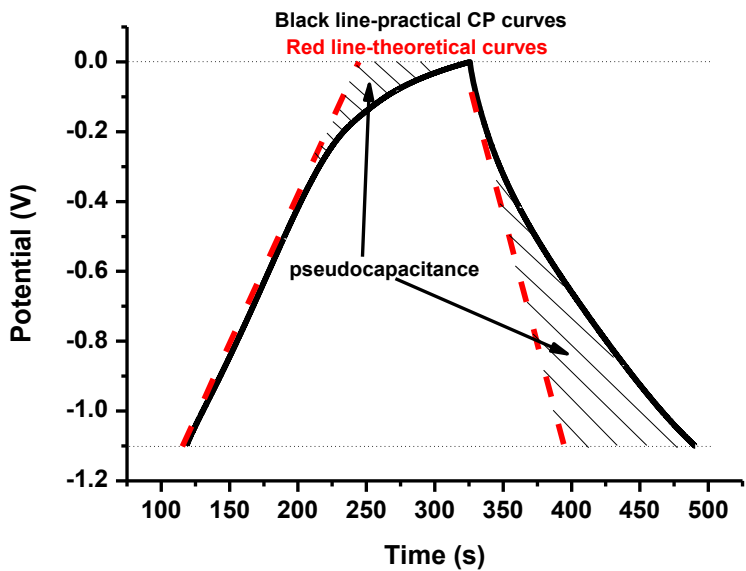

Figure 4. The CP curves of practical and theoretical EDLC electrodes.

Although this method assumes that the $i_p$ is a constant, it is more simple and effective than others. This assumption is reasonable in the case which has a low proportion of $i_p$.

*Metal-based pseudocapacitor electrodes* Because the method in Figure 2 fails in separating current in CV curves of pseudocapacitance (PC), we have to find another simple and effective method following the steps below:

1. We still need Eq.(6);

2. At several points, over non-peak potential range, by equating the slopes (not intercepts) of the $i_{total} - v$ straight-lines, we can obtain the values of $C$ in Eq.(6);

3. The $C$ values from step 2 is then multiplied by $v$, we get $i_d$ at different rates, over non-peak potential range;

4. $i_d$ values are not calculated separately over peak potential range. Finally, separated EDLC current curve can be obtained by connecting all data points from step 3, as shown in Figure 5.

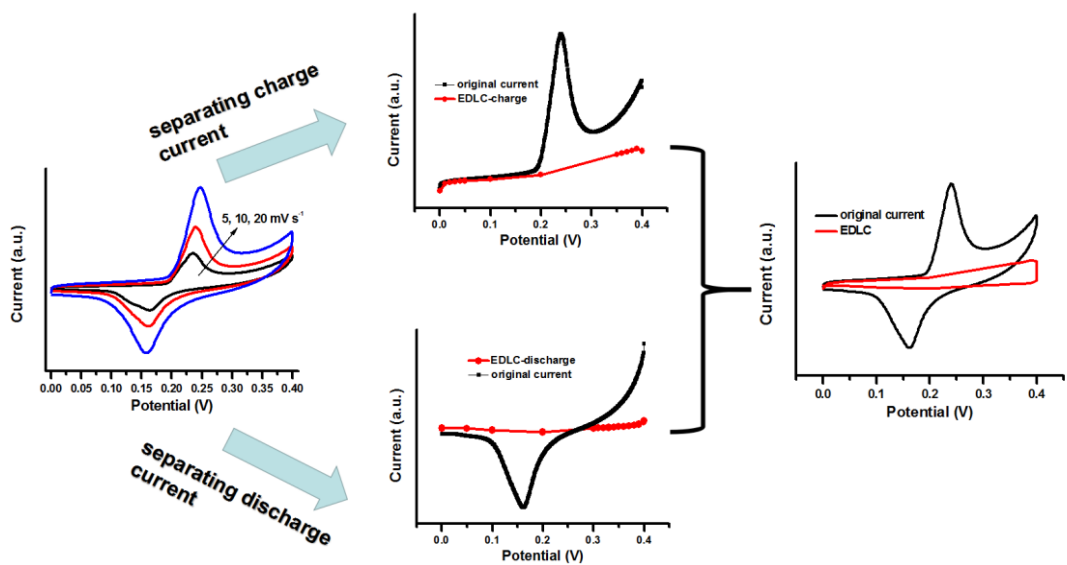

Figure 5. Flow-process diagram of separating current in CV curve of PC.

In Figure 5, many CV curves at different rates must be firstly measured as Figure 3 did. It pays special attention to the influence of the resistances of the solution on peak potential. Therefore, we cannot calculate $i_d$ over peak potential range directly. Moreover, the sweep rates should be not more than 20 mV s$^{-1}$ (10 mV s$^{-1}$ would be better).

An enlarged Figure 5 at low potentials, when separating charge current, is shown in Figure 6.

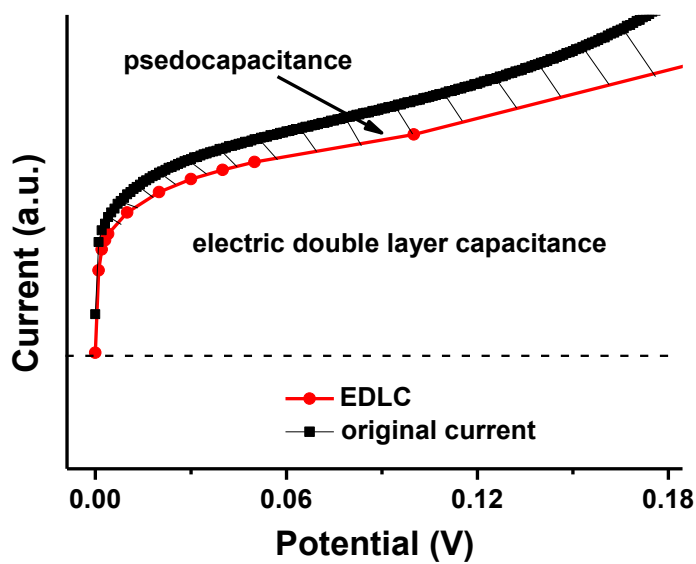

Figure 6. Enlarged Figure 5 at low potentials, when separating charge current.

In practice, the current increases at the very start of many CV curves of PC. It can be seen from Figure 6 that this phenomena occurs because of the charging current for EDLC. Due to the influence of the resistances of the solution on peak potential, Eq.(6) is invalid for separating current over peak potential range. We can only rely on non-peak potential points, even so, this method works very well. In addition, the method eliminates the effect of sweep rate on $i_d$ and $i_p$.

The above discussion is based on a reversible system for ideal supercapacitors, while for other irreversible systems, the current is also proportional to the $\sqrt{v}$ in the *i-E* expression.[4] So the method proposed in this paper can also be used in irreversible systems. And for quasi-reversible system, as the current is not proportional to the $\sqrt{v}$ in the *i-E* expression, only when the sweeping rate is restricted to be enough slow could the current be separated with this method for the quasi-reversible reaction can be regarded as reversible reaction.

In the end, one thing we need to mention here is that the pseudocapacitive electrode materials, such as $MnO_2$ and $RuO_2$, display CV curve typical of that observed for EDLC. However, their energy storage mechanism have not been investigated in detail. We should not treat the electrodes of this type with the above described method before clarifying the mechanism.

## References

[1] L.L. Zhang, X.S. Zhao, Chem. Soc. Rev. 38, 2520, (2009)

[2] P. Simon, Y. Gogotsi, Nat. Mater. 7, 845, (2008)

[3] K.J. Vetter Electrochemical kinetics: theoretical aspects, Elsevier, 2013.

[4] R.S. Nicholson, I. Shain, Anal. Chem. 36, 706, (1964)

[5] E. Gileadi Electrode kinetics. VCH, New York, 1993.